\documentclass{elsart}    

\usepackage{amsmath}
\usepackage{amssymb}
\usepackage{graphicx}
\usepackage{cite}
\usepackage{alltt}
\usepackage{bm} 
\usepackage{url}
\usepackage{longtable}


\newcommand{\nucHe}{\mbox{${}^{4}{\rm He}$}}
\newcommand{\nucLia}{\mbox{${}^{6}{\rm Li}$}}
\newcommand{\nucBea}{\mbox{${}^{7}{\rm Be}$}}
\newcommand{\nucLib}{\mbox{${}^{8}{\rm Li}$}}
\newcommand{\nucBeb}{\mbox{${}^{9}{\rm Be}$}}
\newcommand{\nucBa}{\mbox{${}^{10}{\rm B}$}}
\newcommand{\nucBb}{\mbox{${}^{11}{\rm B}$}}
\newcommand{\nucC}{\mbox{${}^{12}{\rm C}$}}
\newcommand{\nucCb}{\mbox{${}^{13}{\rm C}$}}
\newcommand{\nucN}{\mbox{${}^{14}{\rm N}$}}
\newcommand{\nucNb}{\mbox{${}^{15}{\rm N}$}}
\newcommand{\nucO}{\mbox{${}^{16}{\rm O}$}}

\newcommand{\nucCu}{\mbox{${}^{63}{\rm Cu}$}}
\newcommand{\nucXe}{\mbox{${}^{129}{\rm Xe}$}}
\newcommand{\nucLa}{\mbox{${}^{139}{\rm La}$}}
\newcommand{\nucPb}{\mbox{${}^{208}{\rm Pb}$}}
\newcommand{\nucAu}{\mbox{${}^{197}{\rm Au}$}}
\newcommand{\nucU}{\mbox{${}^{238}{\rm U}$}}

\begin{document}
\begin{frontmatter}

\title{{\tt GLISSANDO 2}: {GL}auber {I}nitial-{S}tate {S}imulation {AND} m{O}re..., ver.~2%
\thanksref{grant}} \thanks[grant]{Supported by Polish National Science Centre, grant DEC-2011/01/D/ST2/00772}

\author[as]{Maciej Rybczy\'nski},
\ead{Maciej.Rybczynski@ujk.edu.pl.pl}
\author[as]{Grzegorz Stefanek},
\ead{Grzegorz.Stefanek@ujk.edu.pl}
\author[as,ifj]{Wojciech Broniowski},
\ead{Wojciech.Broniowski@ujk.edu.pl}
\author[agh,ifj]{Piotr Bo\.zek},
\ead{Piotr.Bozek@ifj.edu.pl}

\address[as]{Institute of Physics, Jan Kochanowski University, 25-406~Kielce, Poland} 
\address[ifj]{The H. Niewodnicza\'nski Institute of Nuclear Physics, Polish Academy of Sciences, 31-342 Cracow, Poland}
\address[agh]{AGH University of Science and Technology, 
    Faculty of Physics and Applied Computer Science, 30-059 Cracow, Poland} 

\begin{abstract}
We present an extended version of {\tt GLISSANDO}, a Monte-Carlo generator for Glauber-like models of the initial 
stage of relativistic heavy-ion collisions. 
The increased functionality of the code incorporates a parametrization of shape of nuclei, including light 
nuclei needed in the NA61 experiment, the nuclear deformation, a possibility of using 
correlated distributions of nucleons in nuclei read from external files, an option of overlaying distributions of produced particles dependent 
on the space-time rapidity, 
the inclusion of the core-corona effect, or the output of the source distributions that can be used in event-by-event hydrodynamics.
Together with other features, such as incorporation of various variants of Glauber models, or 
the implementation of a realistic NN collision profile, the generator 
offers a realistic and practical approach to describe the early phase of the collision in 3+1 dimensions; 
the predictions may later be used in modeling the intermediate evolution phase, e.g., with hydrodynamics.
The software is integrated with the {\tt ROOT} platform.
The supplied scripts compute and plot numerous features of the distributions, such as the 
multiplicity distributions and centrality classes, harmonic asymmetry coefficients and their correlations, forward-backward 
correlations, etc. The code can also be used for the proton-nucleus and deuteron-nucleus collisions. 
\end{abstract}

\date{4 October 2013, sent to Computer Physics Communications}


\end{frontmatter}

\maketitle

\newpage

\noindent{\bf Program summary}

\noindent
{\sl Title of the program:}\-  {\tt GLISSANDO~2}  \hfill  ver. 2.702\\
{\sl Catalog identifier:}\- \\
{\sl Program summary URL:}\\ \- http://www.ujk.edu.pl/homepages/mryb/GLISSANDO/index.html \\
{\sl Program obtainable from:}\\ \- http://www.ujk.edu.pl/homepages/mryb/GLISSANDO/index.html \\
{\sl Licensing provisions:}\- none \\
{\sl Computer:} \- any computer with a {\tt C++} compiler and the {\tt ROOT} environment (optionally with {\tt doxygen}), tested with Intel Xeon X5650, 2.67~GHz, 2~GB RAM\\
{\sl Operating system under which the program has been tested:} \-
{Linux} Ubuntu~7.04-12.04 ({\tt gcc} 4.1.3-4.6.3), Scientific Linux CERN 5.10 ({\tt gcc} 4.1.2), 
{\tt ROOT}~ver.~5.28--5.34/09\\
{\sl Programming language used:} \- {\tt C++} with the {\tt ROOT} libraries \\
{\sl Memory required to execute with typical data:} \- below 120~MB\\
{\sl No. of lines in distributed program, including test data:} \- 3000\\
{\sl No. of bytes in distributed program, including test data and manual:} \- 450~kB\\
{\sl Distribution format:} \- tar.gz\\
{\sl Nature of physical problem:} \- Glauber models of the initial state in relativistic heavy-ion collisions\\
{\sl Method of solution:} \-  Glauber Monte-Carlo simulation of collision events, analyzed with {\tt ROOT}\\
{\sl Restrictions concerning the complexity of the problem:} \- none\\
{\sl Optional software:} \- doxygen~\cite{doxygen:2013}\\
{\sl Typical running time:} \- 80~s/10000~events for the wounded-nucleon model and 100~s/10000~events for the mixed 
model with the $\Gamma$ distribution, minimum-bias Pb+Pb collisions and hard-sphere wounding profile.  
A typical high-statistics ``physics'' run with 500000 events takes about 1 hour. The use of the Gaussian wounding profile increases the time by about a factor of 2. (All times for Intel Xeon X5650, 2.67~GHz, 2~GB RAM)\\

\section{Introduction}

This paper presents an updated and largely enhanced version of the program 
{\tt GLISSANDO} -- {GL}auber {I}nitial-{S}tate {S}imulation {AND} m{O}re..., originally published 
in~\cite{Broniowski:2007nz}.

The popular Glauber~\cite{Glauber:1970wr,Glauber:1987bb,Czyz:1969jg,Miller:2007ri} approach to the early 
phase of relativistic heavy-ion collisions is both physical and 
practical to use in a variety of applications where modelling of the initial phase is needed.  
In this semi-classical approach the individual collisions between 
the nucleons ({\em wounded nucleons}~\cite{Bialas:1976ed,Bialas:2008zza}, possibly admixed with 
{\em binary collisions}~\cite{Kharzeev:2000ph,Back:2001xy}) deposit entropy or energy density 
with a certain distribution in the transverse plane 
and rapidity~\cite{Bialas:2004su,Bialas:2004kt,Gazdzicki:2005rr,Bzdak:2009dr,Bozek:2010bi}. 
The obtained spatial distribution of {\em sources}, which fluctuates event-by-event according to the statistical nature 
of the distributions of nucleons in the colliding nuclei, is usually used 
as input for the intermediate phase of the evolution, typically modelled with relativistic hydrodynamics 
or cascade models (for a review of the heavy-ion 
phenomenology see, e.g.,~\cite{Florkowski:2010zz}). The strength of the sources may fluctuate as well, according to a superimposed 
distribution~\cite{Broniowski:2007nz}. We call this variable {\em relative deposited strength} (RDS).

On the experimental side, the usefulness of the Glauber Monte Carlo simulations comes from the fact that in 
collider experiments one usually determines in this way the dependence of the number of participants on centrality 
\cite{Wang:1991hta,Werner:1988jr,Broniowski:2007nz,Alver:2008aq,Miller:2007ri}. 
On the physics side, the presence of the event-by-event fluctuations in the initial Glauber phase (for recent reviews 
see~\cite{Adare:2012kf,Heinz:2013th}) is a crucial aspect of the approach. 
These geometric fluctuations
\cite{Aguiar:2001ac,Miller:2003kd,Manly:2005zy,Andrade:2006yh,Voloshin:2006gz,Alver:2006wh,Drescher:2006ca,%
Broniowski:2007ft,Voloshin:2007pc,Hama:2009pk,Andrade:2008fa,Broniowski:2009fm,Andrade:2009em,Hirano:2009ah,%
Staig:2010pn,Teaney:2010vd,Qin:2010pf,Nagle:2010zk,Xu:2010du,Lacey:2010av,Werner:2010aa,Qin:2011uw,Bhalerao:2011yg,%
Bhalerao:2011bp,Gardim:2011xv,Qiu:2011hf,Qiu:2011iv,Xu:2011jm,Teaney:2012ke,Jia:2012ma,Hirano:2012kj} 
are carried over to the final distributions of the 
experimentally measured hadrons. They influence the harmonic flow coefficients, in particular generate the odd
components such as the triangular flow~\cite{Alver:2010gr,Alver:2010dn,Petersen:2010cw}, as well as induce the  
correlations between the reaction planes of various harmonic flow 
components~\cite{Adare:2010ux,Nagle:2010zk,Lacey:2010av,Bhalerao:2011yg,Jia:2012ma}.
The output of Glauber Monte Carlo simulations may be used as input for the event-by-event 
hydrodynamics~\cite{Andrade:2006yh,Werner:2009fa,Petersen:2010cw,Holopainen:2010gz,Gardim:2011xv,Bozek:2011if,Schenke:2010rr,Qiu:2011fi,Chaudhuri:2011pa},

Other aspects are also studied theoretically in this approach, such as 
the forward-backward correlations~\cite{Bialas:2004su,Gazdzicki:2005rr,Bzdak:2009dr,Bozek:2010vz,Olszewski:2013qwa}, the 
two-dimensional correlations in relative rapidity and azimuth~\cite{Bozek:2012en}, or the jet 
quenching~\cite{Fries:2010jd,Rodriguez:2010di}.

The applications listed above show that in the active field of studying the initial stage of the relativistic heavy-ion collisions
there is demand for tools implementing the Glauber modeling, which our updated version of {\tt GLISSANDO} tries to satisfy. 
The new features implemented in {\tt GLISSANDO 2} include: 

\begin{itemize}
 \item Parametrization of shape of all typical nuclei, including light nuclei. This is useful in applications for 
       the NA61 experiment, where the mass-number scan will be carried out~\cite{na61url:2013}.
 \item Inclusion of the deformation of the colliding nuclei~\cite{Filip:2007tj,Filip:2009zz,Filip:2010zz}. In particular, the deformation effects 
   are relevant for the collisions involving the deformed Au and U nuclei~\cite{Rybczynski:2012av} recently used at RHIC.
 \item Possibility of using correlated distributions of nucleons in nuclei~\cite{Alvioli:2009ab,Broniowski:2010jd}, which may be 
   read-in from external files prepared earlier with other codes, e.g.,~\cite{Alvioli:page}. Certainly, 
   the two-body correlations are important, as they influence the fluctuations~\cite{Baym:1995cz,Heiselberg:2000fk}.
 \item Generalization of the NN collision profile a shape which interpolates between the step function and a Gaussian 
 profile~\cite{Rybczynski:2011wv}. Such an extension is relevant for the 
 collisions at the LHC energies, allowing to reproduce the measured values of both the total and elastic NN cross sections~\cite{Rybczynski:2013mla}.
 \item Inclusion of the negative binomial overlaid distribution (in addition to the Poissonian and Gamma distributions).
 \item Possibility of overlaying distributions of the produced particles which depend on the space-time rapidity. 
       This feature extends the model into a fully 3+1 dimensional tool.
 \item Inclusion of the core-corona effect~\cite{Hohne:2006ks,Becattini:2008ya,Bozek:2005eu,Werner:2007bf}.
 \item The structure of the {\tt C++} code has been simplified and the organization of the package is restructured.
 \item A {\tt doxygen}-generated \cite{doxygen:2013} reference manual is available, which is useful for those who wish to alter the code for their needs.
\end{itemize}

We recall the original relevant features of the code:

\begin{itemize}

\item Possibility of superimposing a distribution of weights over the distribution of individual sources, reflecting the fact that the elementary
collisions may result in the deposition of a varying amount of the entropy/energy. 

\item The built-in analysis of the shape fluctuations~\cite{Manly:2005zy,Alver:2006wh,Voloshin:2006gz,Broniowski:2007ft}.

\item Evaluation and storage of the two-dimensional density profiles 
to be used ``off-line'' in other analyses, such us the event-by-event initial 
condition for hydrodynamics, jet quenching, {\em etc.} 

\item Output of the event-by-event data that can be used to generate input for hydrodynamics.

\item The code can also be directly used for the proton-nucleus and deuteron-nucleus collisions. The Hulthen distribution 
is used to describe the $NN$ separation in the deuteron.

\item The code uses the standard CERN {\tt ROOT} libraries and data structures. 
 
\end{itemize}

In this paper we only describe the new features of {\tt GLISSANDO 2}, hence the user should also refer to the original 
paper for a more complete description of the physics behind the code and its basic features~\cite{Broniowski:2007nz}. The functionality 
of {\tt GLISSANDO 2} and the format of the input and output files is down-compatible with the original version.

\section{New features in {\tt GLISSANDO 2}}

\subsection{Parametrization of density distributions of light nuclei \label{sec:light}}

For the light nuclei with mass number $3 \leq A\leq 16$, a harmonic oscillator shell model density 
is used~\cite{elton,De Jager:1987qc,Pi:1992ug}:
\begin{eqnarray}
\rho\left(r\right) & = & \frac{4}{\pi^{3/2}C^{3}}\left[1+\frac{A-4}{6}\left(\frac{r}{C}\right)^{2}\right]\exp{(-r^{2}/C^{2})},\nonumber \\
C^{2} & = & \left(\frac{5}{2}-\frac{4}{A}\right)^{-1}\left(\langle r_{ch}^{2}\rangle_{A}-\langle r_{ch}^{2}\rangle_{p}\right),
\end{eqnarray}
where $\langle r_{ch}^{2}\rangle_{A}$ and $\langle r_{ch}^{2}\rangle_{p}=0.7714$~${\rm fm}^{2}$ are the mean squared 
charge radii of the nucleus and the proton, respectively~\cite{angeli_2013}. The values of the 
harmonic oscillator shell model parameter $\langle r_{ch}^{2}\rangle_{A}$ for frequently used light nuclei are 
collected in Table~\ref{Tab:hos_params}.

Since the nucleons are not point-like, the centers of the nucleons cannot be closer than 
particular expulsion distance $d$; this is the usual simple way to introduce the short-range repulsion 
in Glauber Monte Carlo models. The magnitude of $d$ should be of the order of the hard-core repulsion range in the nuclear potential.
The repulsion implemented via an expulsion radius increases somewhat 
the size $R$ of the nucleus and this swelling must be compensated by an appropriate 
shrinkage the parameters of the distribution from which the positions of centers
of nucleons are generated (see Ref.~\cite{Broniowski:2007nz} for a more detailed 
discussion). Accordingly,  when $d>0$, appropriately smaller values of 
the parameter $\langle r_{ch}^{2}\rangle_{A}$ must be used to ensure that the single-particle radial density 
of the simulated nucleus is properly reproduced. These reduced values are given in the third column of Table~\ref{Tab:hos_params}.

\begin{table}
\caption{Harmonic oscillator shell model parameter $\langle r_{ch}^{2}\rangle_{A}$ for several light nuclei~\cite{angeli_2013}.
The values include the case with no NN repulsion ($d=0$) and with the repulsion implemented via expulsion radius of $d=0.9$~fm.
\label{Tab:hos_params}}
\centering
\begin{tabular}{|c|c|c|}\hline
Nucleus & \multicolumn{2}{|c|}{$\langle r_{ch}^{2}\rangle_{A}$~[${\rm fm}^{2}$]} \\ \hline
        &  $d=0$ & $d=0.9$~fm \\ \hline
\nucHe  & $2.81$ & $2.45$ \\
\nucLia & $6.7$ & $6.4$ \\
\nucBea & $7.00$ & $6.69$\\ 
\nucLib & $5.47$ & $5.1$ \\
\nucBeb & $6.35$ & $6.0$\\
\nucBa  & $5.89$ & $5.5$ \\ 
\nucBb  & $5.79$ & $5.36$ \\
\nucC   & $6.10$ & $5.66$\\
\nucCb  & $6.06$ & $5.6$ \\
\nucN   & $6.54$ & $6.08$\\
\nucNb  & $6.79$ & $6.32$ \\ 
\nucO   & $7.29$ & $6.81$ \\\hline
\end{tabular}
\end{table}

\begin{figure}[htb]
\centerline{%
\includegraphics[width=10.0cm]{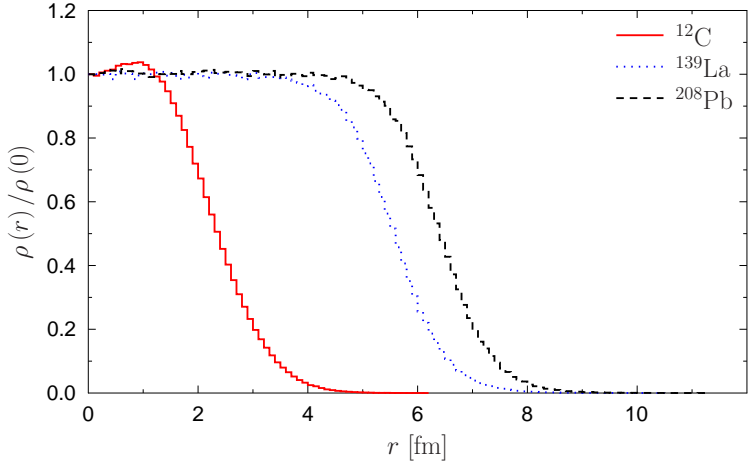}}
\caption{Nuclear density distributions for \nucC~ (harmonic oscillator shell), \nucLa~ and \nucPb~ nuclei.}
\label{Fig:dist}
\end{figure}

We note that the correlated distributions, such as those with $d>0$ are needed only in studies of observables sensitive to correlations, such as, e.g., the multiplicity fluctuations or participant eccentricities of the fireball.

\subsection{Parametrization of density distributions of heavy nuclei \label{sec:heavy}}

For heavy nuclei with $A>16$, the nuclear distributions are well described by the Woods-Saxon 
profiles with the radius and thickness parameters given by
\begin{equation}
R=\left(1.12A^{1/3}-0.86A^{-1/3}\right)~{\rm fm}, \; a=0.54~{\rm fm}.
\label{Eq:RWS_0}
\end{equation}
We recall that Ref.~\cite{Alvioli:2009ab} provide 
distributions of nucleons in nuclei which incorporate  the central repulsive Gaussian two-body correlations between nucleons. 
The one-body Woods-Saxon distributions as 
well as the nucleon-nucleon correlations turn out to be well approximated by the hard-core repulsion 
with $d=0.9$~fm~\cite{Broniowski:2010jd,Rybczynski:2010ad}
and the parametrization~\cite{Broniowski:2007nz}
\begin{eqnarray}
R  =  \left(1.1A^{1/3}-0.656A^{-1/3}\right)~{\rm fm},\; a  =  0.459~{\rm fm},
\label{Eq:RWS_09}
\end{eqnarray}
used in {\tt GLISSANDO 2} when $d=0.9$~fm.

Figure~\ref{Fig:dist} shows the nuclear one-body densities following from the applied parameterizations for a few sample nuclei.

\subsection{Nuclear density distributions with deformation \label{sec:def}}

As originally argued by Filip et al.~\cite{Filip:2007tj,Filip:2009zz,Filip:2010zz}, the nuclear 
deformation plays a relevant role in the ``geometry'' of the collision. In collisions of 
deformed nuclei, the orientation of nuclei relative to each other and to the beam axis influences 
the initial eccentricities and introduces an additional source of the initial fluctuations.  
Since recently the UU collisions were registered at BNL RHIC~\cite{Huang:2012sc,Iordanova:2013jba,FortheSTAR:2013bza}, 
the inclusion of the nuclear deformation in Glauber Monte Carlo simulations is desired~\cite{Rybczynski:2012av}. 
In {\tt GLISSANDO 2}, the spatial distribution of nucleons in colliding heavy nuclei (A$>16$) can be generated from the deformed Woods-Saxon density
\begin{equation}
\rho\left(r\right)=\frac{\rho_{0}}{1+{\rm exp}\left(r-R\left(1+\beta_{2}Y_{20}+\beta_{4}Y_{40}\right)\right)/a}.
\label{Eq:deformed}
\end{equation}
where $\beta_{2}$ and $\beta_{4}$ are the deformation parameters, while 
$Y_{20}$ and $Y_{40}$ are the spherical harmonics. The parameters for the \nucCu~, \nucXe~, \nucAu~, and \nucU~ nuclei, 
which are the nuclides used in the experiments~\cite{Huang:2012sc,Iordanova:2013jba,FortheSTAR:2013bza} at RHIC, 
are listed in Table~\ref{tab:nuclpar}. 

\begin{table}[b]
\caption{The parameters of the Woods-Saxon nuclear density distribution taken from Eqs.~(\ref{Eq:RWS_0}) and (\ref{Eq:RWS_09}), and the deformation coefficients taken from~\cite{Moller:1993ed}.\label{tab:nuclpar}}
\begin{center}
\begin{tabular}{|c|c|c|c|c|c|c|} \hline
 nucleus   & \multicolumn{2}{|c|} {$R$~[fm]} & \multicolumn{2}{|c|} {$a$~[fm]} &   $\beta_{2}$ & $\beta_{4}$\\ \hline
                                              &  $d=0$ & $d=0.9$~fm &  $d=0$ & $d=0.9$~fm \\ \hline
 \nucCu  & 4.24    & 4.21  & 0.54 & 0.459 &   0.162       & -0.006\\ 
 \nucXe  & 5.49    & 5.43  & 0.54 & 0.459 &   0.143       & -0.001 \\ 
 \nucAu  & 6.37    & 6.29  & 0.54 & 0.459 &   -0.13       & -0.03 \\ 
 \nucU   & 6.8    & 6.71   & 0.54 & 0.459 &   0.28        & 0.093  \\  \hline 
\end{tabular}
\end{center}
\end{table} 

\begin{figure}[tb]
\centerline{%
\includegraphics[width=\textwidth]{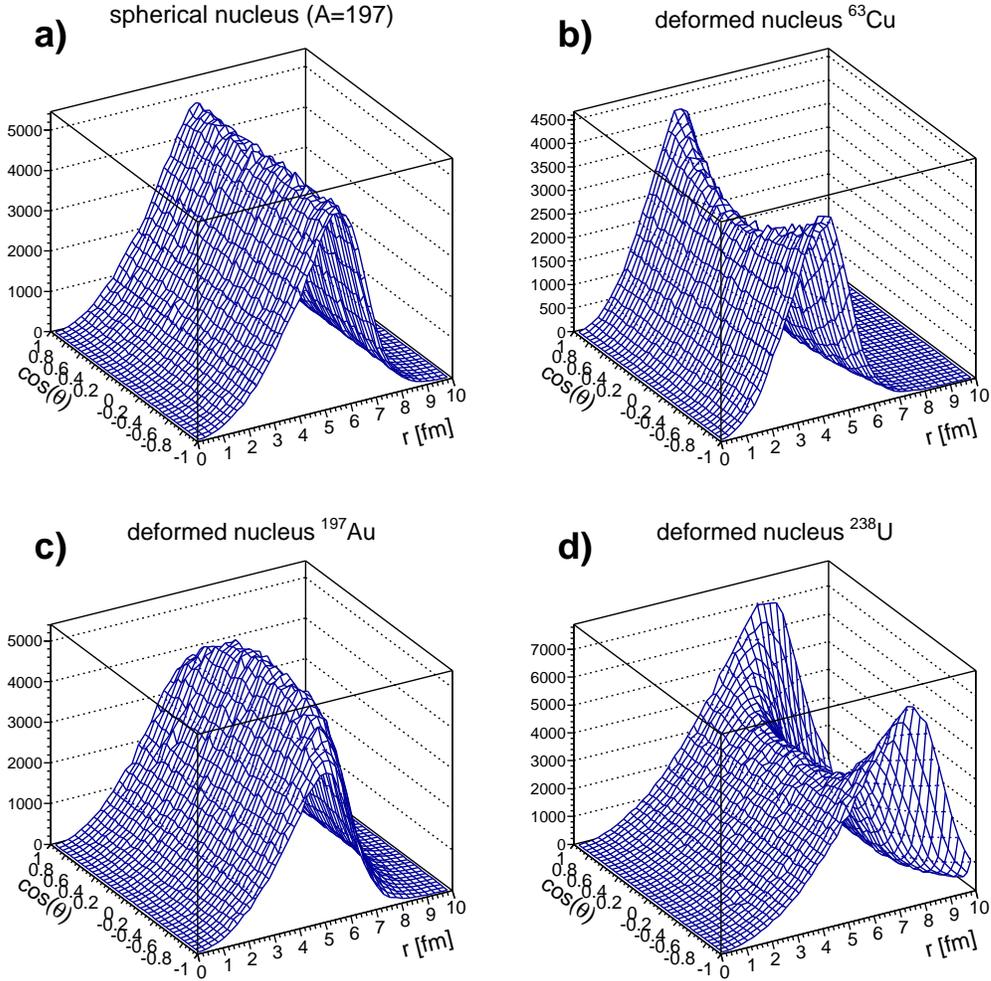}}
\caption{The density profiles of the spherical nucleus with $A=197$~(a) and the deformed nuclei \nucCu~(b), \nucAu~(c), and \nucU~(d).
\label{Fig:profiles}}
\end{figure}

The deformation of these nuclei introduces a significant modification in the shape of the density profiles,
as shown on Fig.~\ref{Fig:profiles}. We compare the density distribution of an (artificially) spherical 
gold nucleus ($A=197$) to the case of the physical \nucAu, exhibiting oblate deformation. We also 
show \nucCu (prolate deformation), and a very strongly deformed \nucU (prolate deformation).    

The deformed nuclear distribution is randomly generated according to Eq.~(\ref{Eq:deformed}) with the symmetry axis aligned
with the beam direction. Before the collision, the nucleus is randomly rotated in three dimensions,
first by the polar angle and later by the azimuthal angle. 

The deformation parameters for the colliding nuclei $A$ and $B$ are called {\tt BETA2A}, 
{\tt BETA4A} and {\tt BETA2B}, {\tt BETA4B}, respectively, and are read from the input file. 
The rotation of nuclei $A$ and $B$ is controlled by four parameters {\tt  ROTA\_PHI}, {\tt ROTA\_THETA} and 
{\tt ROTB\_PHI}, {\tt ROTB\_THETA} respectively. For the default case of random rotation these parameters are set to -1. 
The code also accepts parameters {\tt ROTA\_THETA}, {\tt ROTB\_THETA} equal to the fixed polar angle 
$\theta$ in the range $[0,180]$, and {\tt ROTA\_PHI},  {\tt ROTB\_PHI} equal to the fixed azimuthal angle 
from the range  $[0,360]$. Fixing the rotation angles allows for test simulations with frozen orientations of the 
colliding nuclei. 

\subsection{Collision profile \label{sec:colli}}

Let the total inelastic NN cross section be denoted by $\sigma_{\rm in}$ and the 
corresponding radius parameter $R=\sqrt{\sigma_{\rm in}/\pi}$.
The original version of the code incorporated the popular hard-sphere collision profile (with 
the meaning that the two nucleons collide if their impact parameter is less than $R$), 
\begin{eqnarray}
 p_{\rm HS}(b)=\Theta(R-b),  \label{eq:hs}
\end{eqnarray}
and the Gaussian profile, 
\begin{eqnarray}
 p_{\rm G}(b)=A \exp{\left(-\frac{A b^{2}}{R^2}\right)},   \label{eq:gauss}
\end{eqnarray}
which with $A=0.92$ for the RHIC 
energies led to realistic values of the inelastic and elastic cross sections in the NN collisions~\cite{Rybczynski:2011wv}. At the LHC energies 
a modification of the collision profile is needed to accomplish this goal. We follow Ref.~\cite{Rybczynski:2013mla} and use
\begin{eqnarray}
p_{\rm \Gamma}(b)=G \Gamma\left(\frac{1}{\omega},\frac{Gb^{2}}{R^{2}\omega}\right)/\Gamma\left(\frac{1}{\omega}\right), \label{eq:gamma}
\end{eqnarray}
where $\Gamma\left(z\right)$ and $\Gamma \left(\alpha, z\right)$ denote the Euler
Gamma and incomplete Gamma functions, while $\omega\in\left(0,1\right)$ is a parameter.

\begin{figure}[tb]
\centerline{%
\includegraphics[width=.7\textwidth]{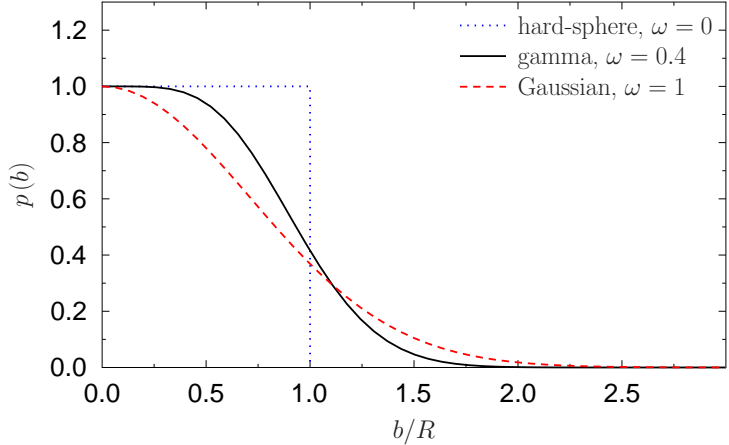}}
\caption{Nucleon-nucleon wounding profile function $p\left(b\right)$ for the hard-sphere, Gaussian and 
Gamma choices. The Gamma profile with parameters $G=1$ and $\omega=0.4$~\cite{Rybczynski:2013mla} approximately
reproduces the TOTEM data~\cite{Antchev:2013iaa} for the elastic differential cross section measured in the 
proton-proton interactions at $\sqrt{s_{NN}}=7$~TeV.}
\label{Fig:profile}
\end{figure}

The profile (\ref{eq:gamma}) smoothly interpolates between (\ref{eq:gauss})(the limit $\omega \to 1 $) and (\ref{eq:hs}) (the limit $\omega \to 0 $).
Importantly, the parametrization (\ref{eq:gamma}) allows to properly reproduce the experimental values  
$\sigma_{\rm in}=73$~mb and $\sigma_{\rm el}=25$~mb~\cite{Antchev:2013iaa}, which is achieved with $G=1$ and 
$\omega=0.4$. In Fig.~\ref{Fig:profile} we show the shapes of the nucleon-nucleon wounding profile functions 
$p\left(b\right)$ for the hard-sphere, Gaussian, and Gamma choices.

The choice of the wounding profile in {\tt GLISSANDO 2} is controlled by the preprocessor directive {\tt \_nnwp\_}.

\subsection{Superposition model \label{sec:sup}}

In the present version of the code we have added the negative binomial as an option for the overlaid distribution. Thus the 
possibilities are: no overlaid distribution ({\tt MODEL=0}), Poisson distribution ({\tt MODEL=1}), gamma distribution ({\tt MODEL=2}), and 
negative binomial distribution ({\tt MODEL=3}). The first three cases are described in the original paper~\cite{Broniowski:2007nz}. The negative binomial distribution generates the discrete weights according to the formula 
\begin{eqnarray}
 g(w;\kappa,k)=\frac{\Gamma(w\kappa+k)}{\Gamma(w\kappa+1)\Gamma(k)}\frac{\left(\frac{\kappa}{k}\right)^{w\kappa}}{\left(1+\frac{\kappa}{k}\right)^{w\kappa+k}}, \;\; w=0,\frac{1}{\kappa},
 \frac{2}{\kappa},\dots. \label{eq:negbin}
\end{eqnarray}
This distribution has $\langle w\rangle=1$ and $\sigma(w)^2=1/{\kappa}+1/{k}$.

The negative binomial distribution can be supplied independently for the wounded nucleons and binary collisions. The parameter $\kappa$ 
is denoted as {\tt Uw} and {\tt Ubin}, respectively, while  $k={\tt Uw}^2/({\tt Vw}-{\tt Uw})$ for wounded nucleons 
or $k={\tt Ubin}^2/({\tt Vbin}-{\tt Ubin})$ for binary collisions. Then $\sigma(w)^2={\tt Uw/Vw^2}$ or  $\sigma(w)^2={\tt Ubin/Vbin}^2$, respectively.

\subsection{Eccentricities \label{sec:ecc}}

\begin{figure}[tb]
\centerline{%
\includegraphics[width=0.7\textwidth]{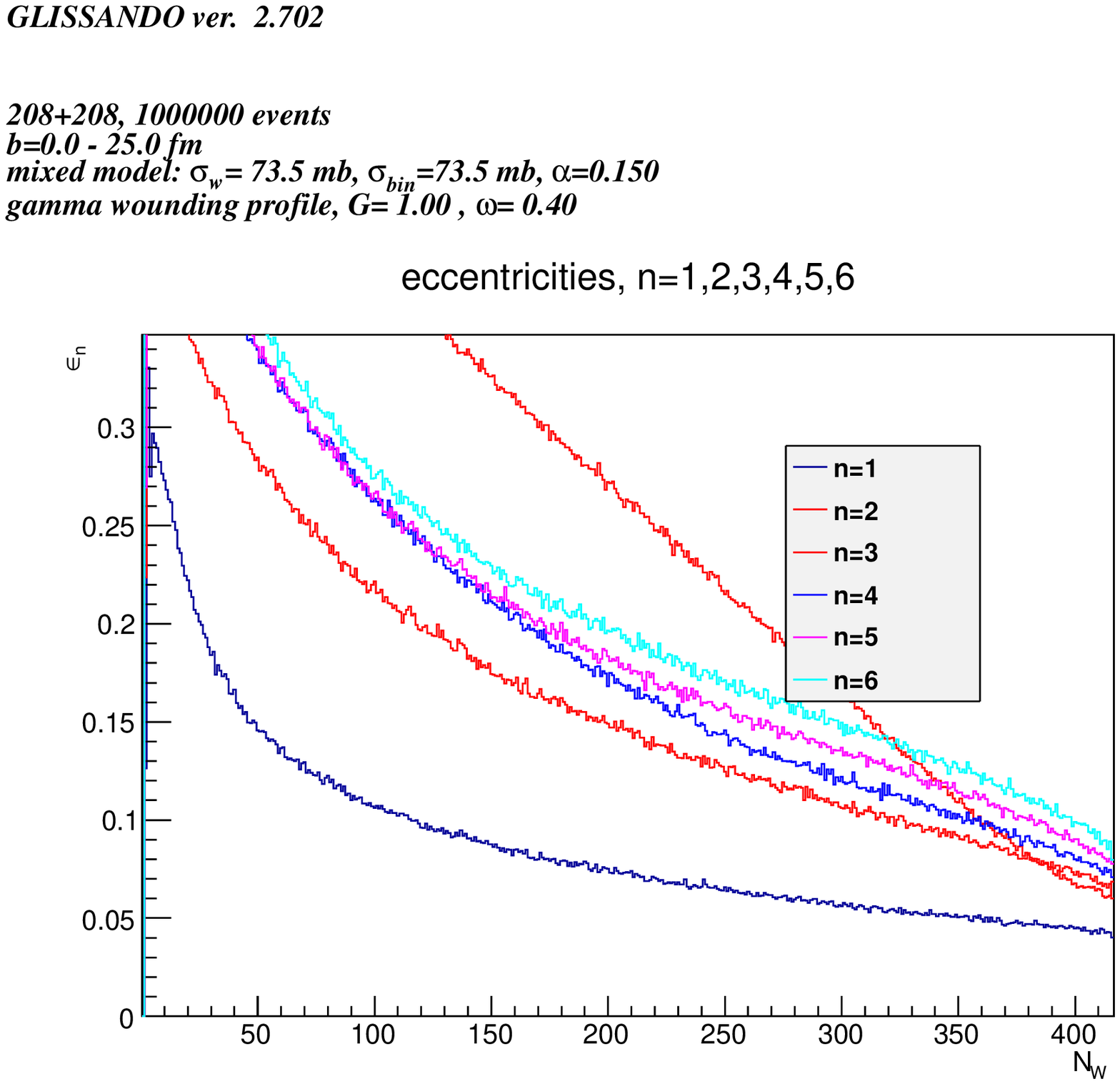}}
\caption{The eccentricities $\epsilon_n^*$ as functions of the number of wounded nucleons for Pb+Pb collisions at the LHC.
\label{Fig:eps}}
\end{figure}

In the present version, all transverse-plane Fourier eccentricity parameters of the created fireball are evaluated as {\em participant eccentricities} (or
variable axes ~\cite{Broniowski:2007nz}) in what became the standard way,
\begin{eqnarray}
\epsilon_n^*=\frac{\langle r^n \cos[n(\phi-\Phi_n)]\rangle} {\langle r^n \rangle}, \; 
\Phi_n={\rm atan2}\left( \frac{\langle r^n \sin(n \phi)\rangle}{\langle r^n \cos(n \phi)\rangle} \right), \nonumber 
\end{eqnarray}
with the exception that for $n=1$ the weight is $r^3$~\cite{Teaney:2010vd}.
An example of a simulation providing the eccentricity parameters is given in Fig.~\ref{Fig:eps}.

\subsection{Core-corona model \label{sec:core}}

\begin{figure}[tb]
\centerline{%
\includegraphics[width=\textwidth]{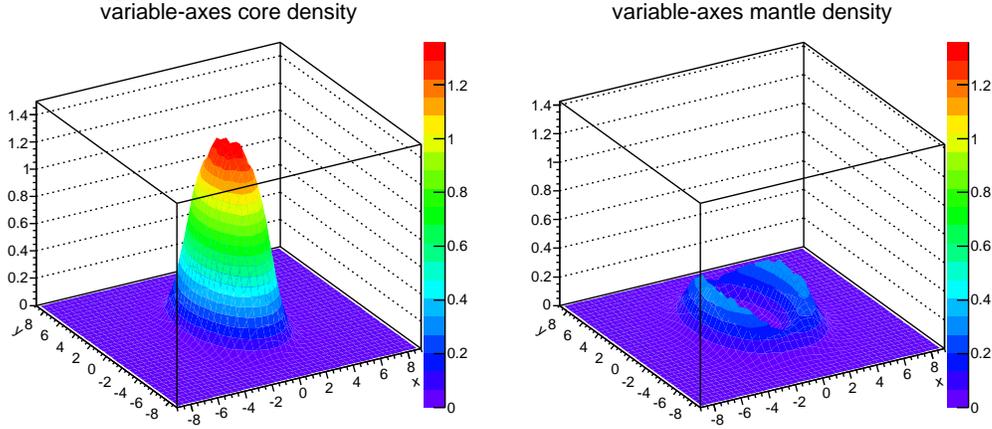}}
\caption{The core and corona distributions.
\label{Fig:core}}
\end{figure}

{\tt GLISSANDO 2} stores the information on how many times a given nucleon interacted with nucleons from the other 
nucleus. This allows for a simple separation of the core (nucleons that interacted more than ones) and corona 
(nucleons that interacted exactly ones)~\cite{Bozek:2005eu,Werner:2007bf,Becattini:2008ya}. 
A sample simulation is presented in Fig.~\ref{Fig:core}.

\subsection{Rapidity distributions \label{sec:rap}}

We implement in the code the following profiles for the space-time rapidity ($\eta_\parallel$) distributions~\cite{Bozek:2010bi}:
\begin{eqnarray}
f(\eta_\parallel)&=&\exp \left ( - \frac{( |\eta_\parallel| -\eta_0)^2}{2 \sigma_\eta^2} \theta(|\eta_\parallel| - \eta_0)\right ), \nonumber \\
f_+(\eta_\parallel)&=&f_F(\eta_\parallel) f(\eta_\parallel), \nonumber \\
f_-(\eta_\parallel)&=&f_F(-\eta_\parallel) f(\eta_\parallel), 
\end{eqnarray}
where
\begin{eqnarray}
f_F(\eta_\parallel)&=&\left \{ \begin{array}{rr} 0, & \eta_\parallel \le -\eta_m 
           \\  \frac{\eta_\parallel+\eta_m}{2 \eta_m}, & -\eta_m < \eta_\parallel < \eta_m \\1, & \eta_m \le \eta_\parallel  \end{array} \right .
\label{eq:F}
\end{eqnarray}
The functions $f_\pm$ are used for the forward (+) and backward (-) moving wounded nucleons, while $f$ is used for the 
binary collisions. 

We adopt a mechanism where a number of sources are generated from each wounded nucleon or binary collision according to the above 
probability distributions. The input parameter {\tt NUMRAP} controls the number of sources, which is equal to 
{\tt NUMRAP*w[i]}, where {\tt w[i]} denotes the weight.

\begin{figure}[tb]
\centerline{%
\includegraphics[width=\textwidth]{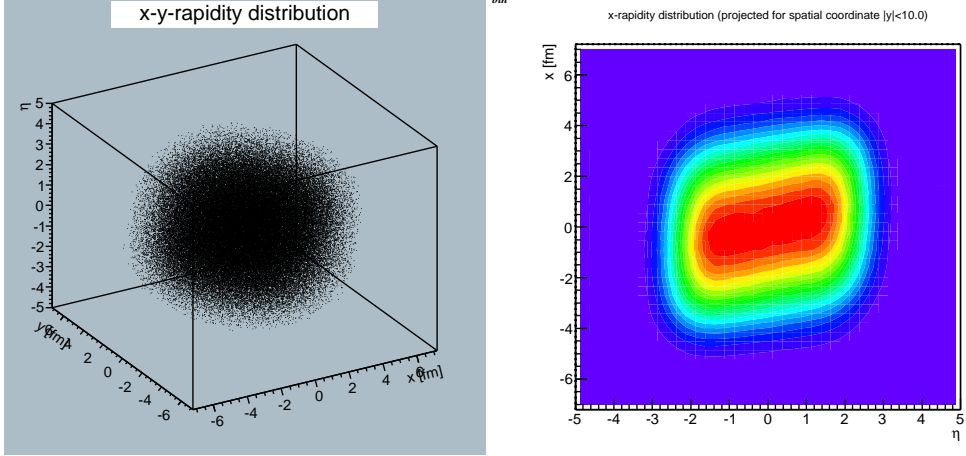}}
\caption{Distributions involving pseudorapidity for Au+Au collisions at RHIC.
\label{Fig:tilt}}
\end{figure}

The following values of the parameters, implemented in the code as {\tt ETA0}, {\tt ETAM}, and {\\tt SIGETA} describe the RHIC data after 
the hydrodynamic evolution~\cite{Bozek:2010bi}:
\begin{eqnarray}
\eta_0=1, \; \eta_m=3.36, \; \sigma_\eta=1.3.
\label{eq:para}
\end{eqnarray}
The emission profiles (\ref{eq:F} were used by one of us (PB)
in Ref.~\cite{Bozek:2010bi} to describe successfully the pseudorapidity spectra and the directed flow in Au+Au collisions at RHIC.
A physical motivation for these ``triangular'' parametrizations has been 
given in~\cite{Bialas:2004su,Nouicer:2004ke,Gazdzicki:2005rr,Bzdak:2009dr,Bozek:2010bi}.
The form (\ref{eq:F}) results in a tilted distribution in the transverse coordinate-spatial pseudorapidity space. 
This is demonstrated in Fig.~\ref{Fig:tilt}.

For the LHC energies, the proper values of the parameters describing the longitudinal distribution of sources can be 
provided after the experimental results for the pseudorapidity spectra become available.

\section{Installation and running \label{sec:install}}

It is necessary to have the {\tt ROOT} package~\cite{root} installed.
After obtaining the {\tt GLISSANDO} distribution, the user should simply run

\begin{verbatim}
make
\end{verbatim}

\noindent which creates the binary file {\tt glissando2}. 

In order to optionally recreate the reference manual, {\tt doxygen}~\cite{doxygen:2013} should be installed first.
and then the command 

\begin{verbatim}
make cleandoc
make doc
\end{verbatim}

\noindent should be executed. This needs to be done only when the user wishes to regenerate the latex reference 
manual and/or create its html version. The configuration is controlled by the supplied 
{\tt Doxyfile}. The original reference manual is provided 
in the distribution in the pdf format as {\tt /doc/latex/refman.pdf}

After the installation, an instructive presentation of the capabilities of the present version
of the code can be carried out with the shell script

\begin{verbatim}
./run.sh
\end{verbatim}

\noindent The created {\tt eps} files are by default displayed with {\tt ghostview}, which should be installed prior to the run. 
Alternatively, the user may edit the file {\tt run.sh} and replace {\tt gv} with his favorite postscript viewer. 
The presentation in {\tt run.sh} goes over typical applications of the code, including the new features, and gives the user a warm-up
before making his own simulations. One can also run the script without prompts by executing

\begin{verbatim}
./run.sh < one.dat
\end{verbatim}

\begin{table}[tb]
\caption{A typical output to the console from {\tt GLISSANDO2}, generated with 
{\tt ./glissando2 input/input\_p\_Pb.dat} (case of central p+Pb collisions
at the LHC). \label{tab:out}}
\begin{center}
{\small
\begin{verbatim}
*********************************************************
GLISSANDO 2 ver. 2.7xx
ver. 2: http://arxiv.org.abs/xx13.xxxx
ver. 1: Computer Physics Communications 180(2009)69, http://arxiv.org.abs/0710.5731
and Phys. Rev. C81(2010)064909 for implementation of the NN correlations
(tested with ROOT ver. 5.28--5.34)
**********************************************************
Simulation of nucleus-nucleus collisions in Glauber models
----------------------------------------------------------
parameters reset from default in input/input_p_Pb.dat :
EVENTS	30000
NUMA	1
NUMB	208
RWSB	-1
ALPHA	0
SNN	73
W0	15
BMAX	7
Woods-Saxon parameters: RWSB=6.40677fm, AWSB=0.459fm (see the paper)

generates Root output file output/glissando.root
random seed: 3045191687, number of events: 30000
1+208, RB=6.40677fm, aB=0.459fm, dB=0.9fm

wounded nucleon model: sig_w=73mb
   (binary collisions not counted)
Gaussian NN collision profile, Ga=0.92
rank of rotation corresponds to the rank of the given Fourier moment
power of transverse radius in eccentricities = rank (see the paper)
window: b_min=0fm, b_max=7fm, Nw_min=15, Nw_max=1000

event: 30000     (100%)              

Some quantities for the specified b, N_w, and RDS window 
(+/- gives the e-by-e standard deviation):
A+B cross section = 306.405mb
efficiency (accepted/all) = 19.9045%
N_w = 17.2806+/-2.21813
relative deposited strength (RDS) = 8.64028+/-1.10906

participant eccentricities:
eps_1 = 0.192552+/-0.107269
eps_2 = 0.305913+/-0.149797
eps_3 = 0.366467+/-0.171138
eps_4 = 0.434756+/-0.195091

Finish: Sat Sep 14 13:40:08 2013
(0h:1m:35s)
\end{verbatim}
}
\end{center}
\end{table}

For simulations of the $A$+$B$ collisions the running command has the syntax

\begin{verbatim}
./glissando2 [input_file] [output_file]
\end{verbatim}

\noindent $A$ and $B$ mean here any nucleus, including the deuteron and the proton.
When the input and output file-name arguments are absent, their default values are 
\begin{verbatim}
input.dat - default input
glissando.root - default output
\end{verbatim}
Typical input files are also provided with the distribution. The input 
parameters and their defaults are described in Appendix~\ref{sec:input}.
Thus we may simply type {\tt ./glissando2} for the basic run.

\section{Customization}

\subsection{Makefile}

The {\tt Makefile} contains instructions for compilation and linking. The user may modify the 
line with the preprocessor options, which control the running mode of the code. The default, needed for the most typical 
simulations, is
{\small
\begin{verbatim}
PREPROCESS := -D_nnwp_=1 -D_files_=0 -D_profile_=0 
                  -D_weight_=0 -D_rapidity_=0 -D_evout_=0
\end{verbatim}
}
The meaning of the parameters is as follows: 
{\small
\begin{verbatim}
 _nnwp_    =2  - use the Gamma wounding profile
           =1  - use Gaussian wounding profile (more realistic),              
           =0  - use the hard-sphere profile
 _files_   =1  - read the nuclear distributions from external files,          
           =0  - generate nuclear distributions randomly
 _profile_ =1  - generate the nucleon profile and NN correlation data,        
           =0  - do not
 _weight_  =1  - generate the NN collision profiles and the RDS ditributions, 
           =0  - do not 
 _rapidity_=1  - generate the data for the rapidity distributions,            
           =0  - do not
 _evout_   =1  - generate event-by-event data
           =0  - do not
\end{verbatim}
}
Instead of modifying the file, the user may run, for instance
\begin{verbatim}
make clean
make 'PREPROCESS = -D_nnwp_=1 -D_rapidity_=1'
\end{verbatim}
to produce the binary code with the Gaussian wounding profile generating the rapidity distributions.

Another functionality is the storage of the current version of the package,
\begin{verbatim}
make package
\end{verbatim}
as well as file cleaning options:
\begin{verbatim}
make clean
make cleandoc
make cleanoutput
\end{verbatim}

\subsection{Input}

The input file is a standard ASCII file. Every line contains the name of the parameter separated with space from its value.  
When the parameter is missing from the input file, or a line containing it is connected out with the \# symbol, a default 
value supplied in the code is used. See Appendix~\ref{sec:input} for details.

\subsection{Output}

\noindent 
A typical output to the console is shown in Table~\ref{tab:out}.
The subsequent self-explanatory lines give the info on the input: the version of the code, initial time, 
name of the input file used and the
values of parameters reset from the default, the seed for the {\tt ROOT} 
random-number generator, the requested number of events, the mass numbers
of nuclei (with 1 corresponding to the proton and 2 to the deuteron), the 
Woods-Saxon parameters, the deformation parameters, the expulsion distance, the type and parameters of the model (wounded, binary, 
mixed, hot-spot), the window in the impact parameter, the number of wounded nucleons or the value of RDS, 
the dispersion parameters for the location of sources, and the live counter for events. The final output consists of the 
total nucleus-nucleus cross section in the given window, $\sigma_{AB}$, the equivalent hard-sphere radius 
defined as $\sqrt{\sigma_{AB}/\pi}/2$, and the efficiency parameter, denoting the ratio of events where the nuclei collided to all 
the Monte-Carlo generated events. Next come averages of basic quantities with their standard deviations: the number of the wounded nucleons, 
binary collisions, RDS, and eccentricity parameters. Finally, the execution time is printed.

The results of the simulation are stored in the {\tt ROOT} output file. To see the physical results, the user should enter 
the {\tt ROOT} environment
\begin{verbatim}
root
\end{verbatim}
\noindent
and execute one of the supplied  scripts. 
An alternative method of executing the scripts is provided in the example shell {\tt run.sh}, for instance one can execute
\begin{verbatim}
./glissando2 input/input_S_Pb.dat output/SPb.root
cd output
root -b -l -q -x "../macro/epsilon.C(\"SPb.root\")"
\end{verbatim}

\subsection{Reading external distributions}

The external files with the distributions have the format

\begin{verbatim}
x   y   z   k
\end{verbatim}

\noindent where x, y, and z denote the Cartesian coordinates of the centers on nucleons in fm, while $k=0$ for neutrons and $k=1$ for 
protons. The file must contain $A*n$ such lines, where $A$ is the mass number of the nucleus, and $n$ is the number of configurations. 
The files with the correlated distributions of Alvioli et al.~\cite{Alvioli:2009ab} for $^{16}$O, $^{14}$Ca, and $^{208}$Pb can be obtained from 
{\tt http://www.phys.psu.edu/∼malvioli/eventgenerator/}. The user must then create from the stored files one big file, for instance running 

\begin{verbatim}
cat o16-1.dat o16-2.dat o16-3.dat [more files] > o16.dat
\end{verbatim}

\noindent The file {\tt o16.dat}  must be placed in the relative subdirectory {\tt nucl}.
To use the external files the code must be compiled through

\begin{verbatim}
make 'PREPROCESS = -D_files_=1'
\end{verbatim}

\noindent and executed as 

{\small
\begin{verbatim}
./glissando2 [input_file] [output_file] [nucleus_A_file] [nucleus_B_file] 
\end{verbatim}
}

\noindent If the syntax 

\begin{verbatim}
./glissando2 [input_file] [output_file] [nucleus_A_file]
\end{verbatim}

\noindent is used, then the distribution of the nucleons in nucleus A is read from the file, while for the nucleus B it is generated randomly. This syntax should
also be used for the collisions of nucleus A with the proton or deuteron.

\begin{figure}[tb]
\centerline{%
\includegraphics[width=0.7\textwidth]{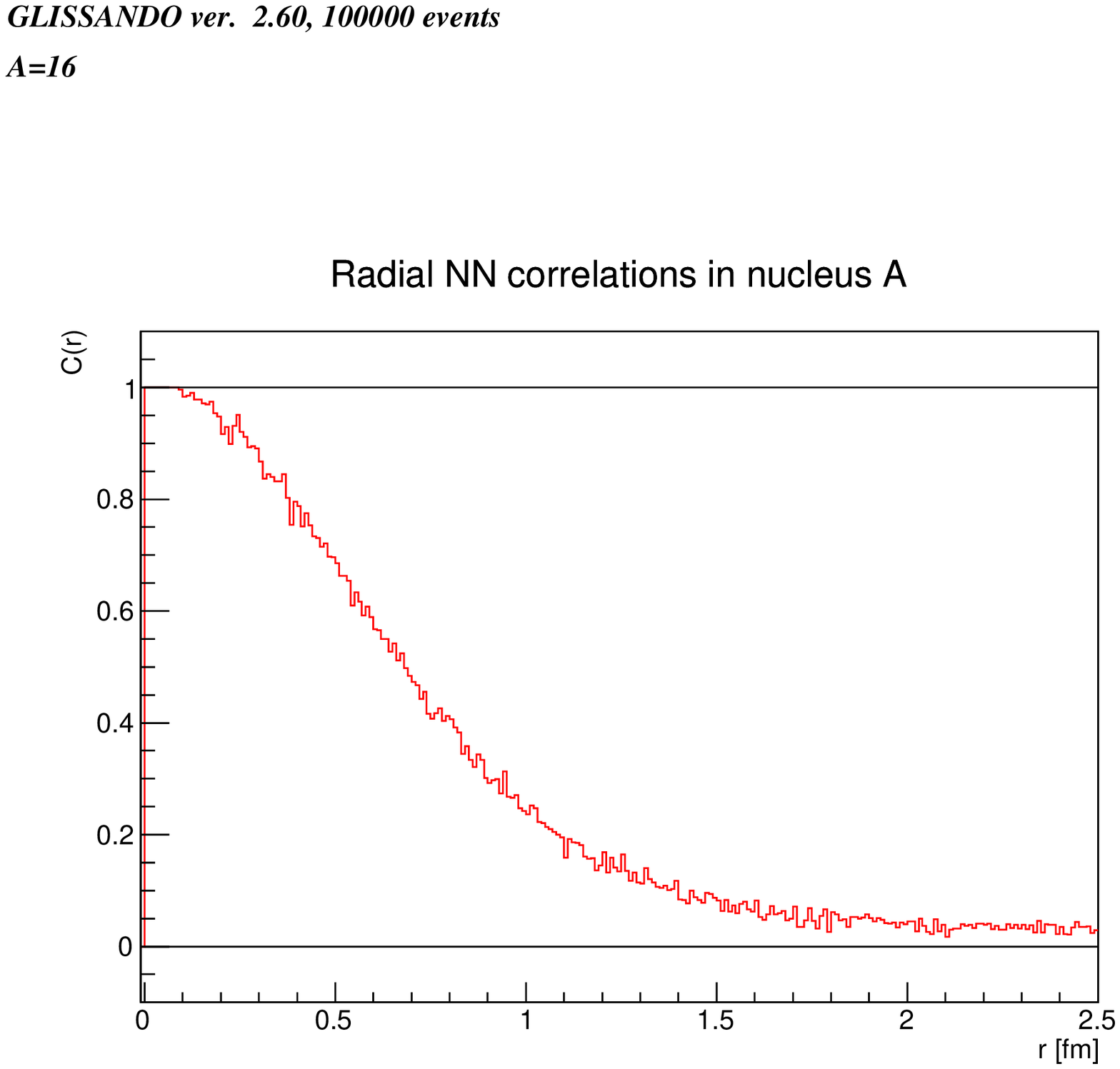}}
\caption{Two-body NN correlations in oxygen 16, generated for the distributions in files downloaded from~\cite{Alvioli:page}.
\label{Fig:co}}
\end{figure}

\subsection{Generating output for hydrodynamics}

The pre-compiler flag {\tt \_evout\_=1} switches the event-by-event output to an external file, storing the position 
of sources and other information in each event to the output file {\tt output/ebye.out}. Its content consists of the blocks 
\begin{verbatim}
n 
\end{verbatim}
followed with $n$ lines of the format
\begin{verbatim}
x  y  z  c  w 
\end{verbatim}
\noindent where {\tt n} is the number of sources in a given collision, 
{\tt x} and {\tt y} are the transverse coordinates in fm, $|c|$ indicates how many times a wounded nucleon collided, with 
positive (negative) {\tt c} corresponding to nucleus A (B), while $c=0$ indicates the binary collisions. The last entry is the weight {\tt w} (RDS).
The number of blocks equals to the number of events.

For the event-averaged densities the user may instead use the script {\tt hydro.C}

\subsection{Fixing centrality cuts \label{sec:cencut}}

Most potential users of the code will run it with their preferred values of 
parameters and typically with division in centrality classes. To carry out the analysis for a given centrality class, a two-step procedure is needed. 
First, a minimum-bias calculation must be done, with no (or broad-range) values for the {\tt W0}, {\tt W1}, {\tt RDS0}, and {\tt RDS1} parameters, as well as {\tt BMIN=0} and {\tt BMAX} set to 
a value approximately equal to twice the sum of the radii of the two colliding nuclei. 
Next, the {\tt macro/centrality2.C} script must be run in root. The values of {\\tt W0}, {\tt W1}, or {\tt RDS0}, {\tt RDS1} determining the centrality classes can be read off from the 
generated file {\tt output/centrality2.dat}. Then the input file must be modified with the proper values for {\tt W0}, {\tt W1} supplied (if centrality is to be determined by the number of the wounded nucleons), or {\tt RDS0}, {\tt RDS1} (if centrality is to be determined by the relative deposited strength RDS).

\section{Structure of the code}

For those readers who may wish to modify the code, we very briefly describe the structure of {\tt GLISSANDO 2}.

As both the nuclei and the Glauber sources created in the collision constitute
certain spatial distributions of points (with weights), we have defined a general class {\tt distr} to create, store, and manipulate such 
distributions (file {\tt build/include/distib.h}). Translations, rotations, evaluation of 
harmonic coefficients, etc., are function members of this class. 
A derivative class {\tt nucleus} is used to create the nuclear distributions, which may either
be generated randomly with a specified distributions, or 
read from prepared earlier external files.

The class {\tt collision} and its derivative {\tt collision\_rap} (file {\tt collision.h}) execute 
the collision of the two nuclei for the case without and with the 
rapidity distributions, respectively. A specific model of the collision and generation of sources is implemented through the 
parameters in the input file.
The result is a spatial distribution of sources with specified weights (RDS).

The auxiliary classes {\tt counter}, {\tt counter2}, and {\tt counter\_2D} (file {\tt counter.h}) 
create useful counters to store and evaluate 
statistical properties such as the mean, variance, etc., of various physical random variables.

Finally, file {\tt build/include/functions2.h} contains definitions of the functions 
(the Woods-Saxon and deformed Woods-Saxon distributions, the Hulthen function, etc.), 
some statistical distributions, structures for the input, initialization of histograms, 
and other technical elements. 

More details concerning the structure of the code are to be found in the supplied reference manual created with {\tt doxygen}. 

\section{Summary}

We have described an extended  version of {\tt GLISSANDO}, hoping it will continue to be a useful tool for the 
heavy-ion community. Moreover, the simplified object-oriented structure of the code, together with the 
technical reference manual, should make it simple to 
tailor to particular needs in future applications. The authors welcome all comments, suggestions, and questions from the users.

\appendix

\section{Contents of the package}

The files included in the {\tt GLISSANDO 2} distribution are listed in Table~\ref{tab:package}.

\begin{table*}[tb]
\caption{The contents of {\tt GLISSANDO 2} package: file names and their descriptions. \label{tab:package} }
\begin{tabular}{lp{4in}}
\hline
file name & description \\
\hline
\vspace{-2mm}
README & basic instructions \\\vspace{-2mm}
Makefile & makefile for {\tt glissando2}\\\vspace{-2mm}
version & stores the minor version number \\\vspace{-2mm}
Doxyfile & configuration for {\tt doxygen} \\\vspace{-2mm}
run.sh & shell script displaying the possibilities \\\vspace{-2mm}
one.dat & auxiliary file \\\vspace{-2mm}
build/src/glissando2.cxx & the {\tt GLISSANDO 2} source file \\\vspace{-2mm}
build/include/functions2.h & the function library \\\vspace{-2mm}
build/include/collisions.h   & the collisions library \\\vspace{-2mm}
build/include/distrib.h   & the distributions library \\\vspace{-2mm}
build/include/counter.h   & the counter library \\\vspace{-2mm}
addons/interpolation.cxx  & source for interpolation code \\\vspace{-2mm}
addons/interpolation.mk   & make file for interpolation code \\\vspace{-2mm}
addons/retrieve.cxx       & template code for retrieving info from the full event tree \\\vspace{-2mm} 
addons/retrieve.mk        & makefile for retrieve \\\vspace{-2mm} 
input/input*.dat          & input files for various collisions \\\vspace{-2mm}
macro/angles.C            & script generating the plot of the correlation between principal axes in the forward and backward rapidities\\\vspace{-2mm}
macro/centrality2.C       & script generating centrality classes \\\vspace{-2mm}
macro/core\_mantle.C      & script generating the core and corona distributions \\\vspace{-2mm}
macro/corr.C              & script generating the NN correlation plot \\\vspace{-2mm}
macro/density.C           & script generating the distributions \\\vspace{-2mm}
macro/dxdy.C              & script for center-of-mass coordinates vs. $N_w$\\\vspace{-2mm}
macro/epsilon.C                 & script for eccentricity vs $N_w$\\\vspace{-2mm}
macro/epsilon\_b.C              & script for eccentricity vs $b$\\\vspace{-2mm}
macro/epsilon\_c.C              & script for eccentricity vs centrality\\\vspace{-2mm}
macro/fitr.C              & script displaying and fitting the nuclear density profile \\\vspace{-2mm}
macro/fourier.C           & script generating first few harnonic components of the distributions  \\\vspace{-2mm}
macro/hydro.C             & script generating input grid for hydrodynamic calculations \\\vspace{-2mm}
macro/info.C              & script giving information on the stored output file\\\vspace{-2mm}
macro/label.C             & script generating the label used in plots \\\vspace{-2mm}
macro/mult.C              & script for multiplicity fluctuations \\\vspace{-2mm}
macro/overlay.C           & script examining the overlaid distributions \\\vspace{-2mm}
macro/profile2.C                & script for Fourier profiles\\\vspace{-2mm}
macro/profile2\_deformation\_*.C   & scripts for r-cos($\theta$) profiles of deformed nuclei\\\vspace{-2mm}
macro/size.C              & script generating the event-by-event scaled standard deviation of the size parameter\\\vspace{-2mm}
macro/tilted.C             & script generating the tilted initial profile in the x-rapidity space at y=0\\\vspace{-2mm}
macro/wounding\_profile.C   & script generating the wounding and binary-collision profiles\\\vspace{-2mm}
doc/latex/refman.pdf      & the {\tt doxygen} reference manual\\\vspace{-2mm}
\end{tabular}
\end{table*}

\section{Description of input and output \label{sec:input}}

The basic model parameters, collected in Table~\ref{tab:input}, can be supplied in the input file. The sign \# at the 
beginning of the line comments out the line and then the default value of the parameter set in the code is used.
See the sample file {\tt input.dat}. 

The meaning of variables stored in the output {\tt GLISSANDO 2 Root} files is 
explained in Tables~\ref{tab:out1} and \ref{tab:out2}.

\begin{longtable}{lrp{4in}}
\caption{Parameters of the input file. \label{tab:input}}\\
\hline
{name} & 
{default} & 
{description} \\
\hline \vspace{-3mm}
\endfirsthead


\multicolumn{3}{c}{{\tablename} \thetable{} -- Continued} \\[0.5ex]
\hline \hline
{name} & 
{default} & 
{description} \\
\hline \vspace{-3mm}
\endhead


\multicolumn{3}{c}{{Continued on Next Page\ldots}} \\
\endfoot


\hline \hline
\endlastfoot


{ISEED}  & {0}     &  {seed for the random number generator, if 0 a random seed is generated} \\\vspace{-3mm}
{EVENTS} & {50000} &  {number of generated events} \\\vspace{-3mm}
{NBIN}   & {40}    &  {number of bins for histograms in $\rho$, $x$, or $y$} \\\vspace{-3mm}
{FBIN}   & {72}    &  {number of bins for histograms in the azimuthal angle} \\\vspace{-3mm}
{NUMA}   & {208}   &  {mass number of nucleus $A$} \\\vspace{-3mm}
{NUMB}   & {208}   &  {mass number of nucleus $B$} \\\vspace{-3mm}
{RWSA}   & {6.407} &  {Woods-Saxon radius for the distribution of centers, nucleus $A$ [fm] (208Pb with the fix-last method)} \\\vspace{-3mm}
{AWSA}   & {0.459} &  {Woods-Saxon width, nucleus $A$ [fm]}\\\vspace{-3mm}
{BETA2A} & {0.}   &  {deformation parameter $\beta_{2}$, nucleus $A$}\\\vspace{-3mm}
{BETA4A} & {0.}   & {deformation parameter $\beta_{4}$, nucleus $A$}\\\vspace{-3mm}
{ROTA\_THETA} & {-1} & {rotation parameter (angle $\theta$), -1 - random rotation, nucleus $A$} \\\vspace{-3mm} 
{ROTA\_PHI} & {-1} & {rotation parameter (angle $\phi$), -1 - random rotation, nucleus $A$} \\\vspace{-3mm} 
{RWSB}   & {6.407} &  {Woods-Saxon radius for the distribution of centers, nucleus $B$ [fm]}\\\vspace{-3mm}
{AWSB}   & {0.459} &  {Woods-Saxon width, nucleus $B$ [fm]}\\\vspace{-3mm}
{BETA2B} & {0.}   & {deformation parameter $\beta_{2}$, nucleus $B$}\\\vspace{-3mm}
{BETA4B} & {0.0}   & {deformation parameter $\beta_{4}$, nucleus $B$}\\\vspace{-3mm}
{ROTB\_THETA} & {-1} & {rotation parameter (angle $\theta$), -1 - random rotation, nucleus $B$} \\\vspace{-3mm} 
{ROTB\_PHI} & {-1} & {rotation parameter (angle $\phi$), -1 - random rotation, nucleus $B$} \\\vspace{-3mm} 
{RCHA}   & {5.66}  & {harmonic oscillator shell model density mean squared charge radii of nucleus $A$ (12C-nucleus)}\\\vspace{-3mm}
{RCHB}   & {5.66}  & {harmonic oscillator shell model density mean squared charge radii of nucleus $B$ (12C-nucleus)} \\\vspace{-3mm} 
{RCHP}   & {0.7714} & {harmonic oscillator shell model density mean squared charge radii of  proton} \\\vspace{-3mm} 
{WFA}    & {0}     &  {the w parameter of the Fermi distribution, nucleus $A$}\\\vspace{-3mm}
{WFB}    & {0}     &  {the w parameter of the Fermi distribution, nucleus $B$}\\\vspace{-3mm}
{CD}     & {0.9}   &  {closest allowed distance between centers of nucleons [fm]}\\\vspace{-3mm}
{SNN}    & {73.5}  &  {$NN$ ``wounding'' cross section [mb]}\\\vspace{-3mm}
{SBIN}   & {73.5}  &  {$NN$ binary cross section [mb]}\\\vspace{-3mm}
{ALPHA}  & {0.15}  &  {0 - wounded, 1 - binary, 0.145 - LHC@2.76~TeV/nucleon}\\\vspace{-3mm}
{MODEL}  & {0}     &  {0 - constant superimposed weight=1, 1 - Poisson, 2 - Gamma, 3 - Negative Binomial} \\\vspace{-3mm} 
{Uw}     & {2.  }   &  {Poisson, Gamma or NegBin parameter for wounded}\\\vspace{-3mm}
{Ubin}   & {2.}     &  {Poisson, Gamma or NegBin parameter for binary}\\\vspace{-3mm}
{Vw}     & {4.}     &  {Negative binomial variance, wounded nucleons}\\\vspace{-3mm} 
{Vbin}   & {4.}     &  {Negative binomial variance, binary collisions}\\\vspace{-3mm} 
{DW}     & {0.}     &  {dispersion of the location of the source for wounded nucleons [fm]}\\\vspace{-3mm}
{DBIN}   & {0.}     &  {dispersion of the location of the source for binary collisions [fm]}\\\vspace{-3mm}
{WMIN}   & {2}     &  {minimum number of wounded nucleons to record the event}\\\vspace{-3mm}
{W0}     & {2}     &  {minimum allowed number of wounded nucleons}\\\vspace{-3mm}
{W1}     & {1000}  &  {maximum allowed number of wounded nucleons}\\\vspace{-3mm}
{RDS0}   & {0}     &  {minimum allowed RDS}\\\vspace{-3mm}
{RDS1}   & {100000} &  {maximum allowed RDS}\\\vspace{-3mm}
{NNWP}   & {0}     &  {0 - hard-sphere NN wounding profile, 1 - Gaussian NN wounding profile, 2 - Gamma NN wounding profile} \\\vspace{-3mm} 
{GA}     & {0.92}  &  {central value of the Gaussian wounding profile}\\\vspace{-3mm}
{GAMA}   & {1.}     & {central value of the Gamma wounding profile}\\\vspace{-3mm}
{OMEGA}  & {0.4}   & {relative variance of cross-section fluctuations for the Gamma wounding profile} \\\vspace{-3mm} 
{SHIFT}  & {1}     &  {1 - shift the coordinates of the fireball to the c.m. in the fixed-axes case, 0 - do not shift} \\\vspace{-3mm} 
{RET}    & {0}     &  {0 - fix-last algorithm, 1 - return-to-beginning algorithm for nuclear density}\\\vspace{-3mm}
{FULL}   & {0}     &  {1 - provide the full information on events (obsolete), 0 - do not}\\\vspace{-3mm}
{DOBIN}  & {0}     &  {1 - compute the binary collisions also for the case ALPHA=0, 0 - do not  compute the binary collisions for the case ALPHA=0} \\\vspace{-3mm}
{FILES}  & {0}     &  {1 - read distribution from files, 0 - do not} \\\vspace{-3mm}
{NUMRAP} & {10}    & {number of particles per unit weight generated in the whole rapidity range} \\\vspace{-3mm} 
{RAPRANGE} & {5.}   & {range in rapidity}\\\vspace{-3mm}
{ETA0}   & {1.}     & {2*ETA0 is the width of the plateau in $\eta$}\\\vspace{-3mm}
{ETAM}   & {3.36}  & {parameter of the Bialas-Czyz-Bozek model}\\\vspace{-3mm}
{SIGETA} & {1.3}   & {parameter controlling the width of the rapidity distribution} \\\vspace{-3mm} 
{MAXYRAP} & {10.}   & {maximum absolute value of the y coordinate in the x-y-rapidity histogram} \\\vspace{-3mm} 
{FBRAP} & {2.5}    & {forward rapidity for the forward-backward analysis (backward rapidity = - FBRAP)} \\\vspace{-3mm} 
{ARANK}  & {2}     & {rank of the Fourier moment for the forward-backward analysis}\\\vspace{-3mm}
{PP}  & {-1}    & {power of the transverse radius in the Fourier moments} \\\vspace{-3mm}
{RO}  & {0}    & {rank of the rotation axes (0 - rotation rank = rank of the Fourier moment)} \\\vspace{-3mm} 
{PI}     & {$4.\arctan(1.)$} &  {the number $\pi$}\\\vspace{-3mm}
{BMIN}   & {0.}    &  {minimum impact parameter [fm]}\\\vspace{-3mm}
{BMAX}   & {25.}   &  {maximum impact parameter [fm]}\\
{BTOT}   & { } &  {range parameter for histograms [fm]}\\

\hline

\end{longtable}

\begin{table*}[tb]
\caption{Some of histograms stored in the output {\tt ROOT} file. $\langle . \rangle$ denotes the mean and var the variance 
of the specified quantity. \label{tab:out1}}
\begin{tabular}{ll}
\hline
\vspace{-3mm}

xyhistr & variable-axes density in the $x-y$ variables\\ \vspace{-3mm}

c0rhist & variable-axes density in the $\rho-\phi$ variables (not normalized)\\\vspace{-3mm}

c0rhp & $f^*_0(\rho)$  [see Eq.~(22-23) in~\cite{Broniowski:2007nz} for the notation below] \\ \vspace{-3mm}
c2rhp & $f^*_2(\rho)$ 
\\ \vspace{-3mm}
c4rhp & $f^*_4(\rho)$ 
\\ \vspace{-3mm}
c6rhp & $f^*_6(\rho)$ 
\\ \vspace{-3mm}
s1rhp & $g^*_1(\rho)$ 
\\ \vspace{-3mm}
s3rhp & $g^*_3(\rho)$ 
\\ \vspace{-3mm}
nx   &    $\langle x \rangle$ [fm] vs. $N_w$\\ \vspace{-3mm}
nx2  &    ${\rm var}(x)$ [fm]$^2$ vs. $N_w$\\ \vspace{-3mm}
ny   &    $\langle y \rangle$ [fm] vs. $N_w$\\ \vspace{-3mm}
ny2  &    ${\rm var}(y)$ [fm]$^2$ vs. $N_w$\\ \vspace{-3mm}
nepsp &   $\langle \epsilon^* \rangle$ vs. $N_w$\\ \vspace{-3mm}
nepsp2 &  ${\rm var}(\epsilon^*)/\langle \epsilon^* \rangle^2$ vs. $N_w$\\ \vspace{-3mm}
nuni   &  event multiplicity vs. $N_w$\\ \vspace{-3mm} 
nepspb &   $\langle \epsilon^* \rangle $ vs. $b$\\ \vspace{-3mm}
nepsp2b &  ${\rm var}(\epsilon^*)/\langle \epsilon^* \rangle^2$ vs. $b$\\ \vspace{-3mm}
nunib   &  event multiplicity vs. $b$\\ \vspace{-3mm}
ntarg  &   $\langle N_w^B \rangle $ vs. $N_w^A$\\ \vspace{-3mm}
ntarg2 &   ${\rm var}(N_w^B)/\langle N_w^B \rangle$ vs. $N_w^A$\\ \vspace{-3mm}
nbinar  &   $\langle N_{\rm bin} \rangle$ vs. $N_w^A$\\ \vspace{-3mm}
nbinar2 &   ${\rm var}(N_{\rm bin})/\langle N_{\rm bin} \rangle$ vs. $N_w^A$\\ \vspace{-3mm}
nwei  &   $\langle RDS \rangle$ vs. $N_w^A$\\ \vspace{-3mm}
nwei2 &   ${\rm var}(RDS)/\langle RDS \rangle$ vs. $N_w^A$\\ \vspace{-3mm}
nuni   &  event multiplicity vs. $N_w^A$\\ \vspace{-3mm} \\
\hline
\end{tabular}
\end{table*}

\begin{table*}[tb]
\caption{Trees and their contents stored in the output {\tt ROOT} file.\label{tab:out2}}
\begin{tabular}{ll}
\hline
TTree param &  all parameters of the calculation \\
\hline
TTree density &  (generated only by {\tt glissando\_profile.exe})\\ \vspace{-3mm} 
r  &   radius of the nucleon in nucleus $A$ [fm]\\
wd &   weight generated by the superposition distribution\\\hline
TTree phys & \\\vspace{-3mm}
sitot & the nucleus-nucleus cross section [mb]\\\vspace{-3mm}
eps\_variable & event-by-event average $\epsilon^\ast$ \\\vspace{-3mm}
sigma\_eps\_variable & event-by-event standard deviation of $\epsilon^\ast$ \\ \vspace{-3mm} \\
\hline
\end{tabular}
%
\begin{tabular}{ll}  
TTree events & \\ \vspace{-3mm}
nwA  & number of wounded nucleons in $A$\\ \vspace{-3mm}
nwB  & number of wounded nucleons in $B$\\ \vspace{-3mm}
nwAB & total number of wounded nucleons\\ \vspace{-3mm}
nbin & number of binary collisions\\ \vspace{-3mm}
npa  & RDS\\  \vspace{-1mm}
b    & impact parameter\\  \vspace{-3mm}
size & weighted average of the distance fom origin (c.m. frame)\\  \vspace{-3mm}
ep1  & $< r^3 \cos( \phi-\phi^*) >/< r^3 >$\\ \vspace{-3mm}
ep   & $< r^2 \cos(2 (\phi-\phi^*)) >/< r^2 >$ \\  \vspace{-3mm}
ep3  & $< r^3 \cos(3 (\phi-\phi^*)) >/< r^3 >$\\ \vspace{-3mm}
ep4  & $< r^4 \cos(4 (\phi-\phi^*)) >/< r^4 >$\\ \vspace{-3mm}
ep5  & $< r^5 \cos(5 (\phi-\phi^*)) >/< r^5 >$\\ \vspace{-3mm}
ep6  & $< r^6 \cos(6 (\phi-\phi^*)) >/< r^6 >$\\ \vspace{-3mm}
phir & the rotation angle $\phi^*$\\ \vspace{-3mm}
phi2\_plus & the rotation angle $\phi^*$, increased rapidity\\ \vspace{-3mm}
phi2\_minus & the rotation angle $\phi^*$, decreased rapidity\\ \vspace{-3mm}
phir3 & the rotation angle $\phi_{3}^*$\\ \vspace{-3mm}
phir4 & the rotation angle $\phi_{4}^*$\\ \vspace{-3mm}
phir5 & the rotation angle $\phi_{5}^*$\\ \vspace{-3mm}
phir6 & the rotation angle $\phi_{6}^*$\\ \vspace{-3mm}
xx   & $x$ c.m. coordinate [fm] (before shifting)\\ \vspace{-3mm}
yy   & $y$ c.m. coordinate [fm] (before shifting)\\ \vspace{-3mm}\\ \hline
\end{tabular}
\end{table*}

\end{document}